\documentclass[
    ,final            
  ]
  {aipproc}

\usepackage{graphicx}
\usepackage{braket}
\usepackage{amssymb}
\usepackage{amsmath}

\layoutstyle{6x9}

\newcommand{\BRA}[1]{{\langle #1 |}}
\newcommand{\KET}[1]{{| #1 \rangle}}

\newcommand{\caL}{{\mathcal L}}
\newcommand{\caR}{{\mathcal R}}

\begin{document}

\title{Exact computation of current cumulants in small Markovian systems}


\author{Marco Baiesi}{
  address={Instituut voor Theoretische Fysica, K.U.Leuven, Belgium.}
}

\author{Christian Maes}{
  address={Instituut voor Theoretische Fysica, K.U.Leuven, Belgium.}
}

\author{Karel Neto\v{c}n\'{y}}{
  address={Institute of Physics AS CR, Prague, Czech Republic.}
}

\begin{abstract}
We describe an algorithm computing the exact value of
the mean current, its variance, and higher order cumulants
for stochastic driven systems.
The method uses a Rayleigh-Schr\"odinger perturbation expansion
of the generating function of the current, and can be extended to
compute covariances of multiple currents.
As an example of application of the method, we give numerical evidence for
a simple relation  [Eq.~(\ref{eq:C4})] between the second and the 
fourth cumulants of the current in a 
symmetric exclusion process.

\end{abstract}

\maketitle


\section{Goal}
In phenomena like charge transport in nano-electromechanical
systems~\cite{rec,but,naz,nov,tom,lev} or in life processes like
molecular motors or like ion-transport through a single membrane
channel, one easily reaches energy scales as low as a few
$k_BT$~\cite{bio,kelvin}. Therefore, the physics and the chemistry
of these small systems must talk about fluctuations, not only
because they are very much present but also because some processes
actually make use of these fluctuations.  Moreover, the
experimental output is often in terms of current cumulants, which
should inform us about important features of the system dynamics.

A central quantity is the fluctuating current $j$, which gives the
time-averaged number of particles or quanta that pass trough a
given surface, and its large deviation function $I(j)$, describing
the shape of its probability distribution $p(j)\simeq \exp(-T\,
I(j))$ for very long time intervals $T$. One is interested in its
full shape, as its tail can contain signatures of interesting
physics. However, the asymptotic estimate of these tails is
problematic as they involve fluctuations that become rare for
$T\to \infty$. We describe a numerical scheme to compute exactly
(up to numerical round-off's) the cumulants of these current
fluctuations in general small Markovian systems. We emphasize that
our algorithm is exact and can be systematically implemented to
produce cumulants of in principle arbitrary order. Practical
limitations such as memory storage make it however most efficient
for the statistical mechanics of systems that are not too large.
Finally, our method is based on general theoretical considerations
and its interpretation involves a relation between current and
traffic (dynamical activity) fluctuations.

Mathematically, our approach uses a type of Rayleigh-Schr\"odinger
perturbation expansion for the largest eigenvalue of a matrix that
is obtained as a perturbation of the original Markov generator;
a full account can be found in~\cite{bmn08}.
It can be seen as a modification and adaption to classical nonequilibrium models, of a technique developed within the framework of full counting statistics for quantum transport~\cite{novc}, see also~\cite{nov,tom,gen,kin,ger,Schm,derc}.
Our approach is somewhat complementary to a recent efficient algorithm for the
(approximate) estimate of the large deviation function~\cite{gia,lec}.

\section{Example model}

To illustrate our focus we consider the well studied
one-dimensional boundary driven simple symmetric exclusion process (SSEP),
for which several rigorous results are
available~\cite{derex,derrida}.
A state is represented by an array
$(\eta_i)$ of $N$ units, which can either be empty ($\eta_i=0$) or
occupied by one particle ($\eta_i=1$). A transition takes place when
a particle moves to a neighboring empty site (with rate $1$ )
or if it enters/exits from one of the end sites,
which are in contact with reservoirs at
different chemical potentials 
($\alpha$ at $i=1$ and $\alpha'$ at $i=N$). 
For example, a transition from
a state $\eta$ to a state $\xi$ due to the entrance of a particle
from the left reservoir ($\eta_1=0 \to \eta_1=1$ and the rest of the
array is unchanged) takes place with rate
$ k(\eta,\xi) = \exp(\alpha/2)$.
Observe that we impose the physical condition of local detailed
balance, meaning that the rates should obey
\begin{equation}\label{ldb}
  k(\eta,\xi) = a(\eta,\xi)\,\exp\Bigl[
\frac{\text{entropy flux}}{2}\Bigr]
\end{equation}
with some symmetric prefactor $a(\eta,\xi)=a(\xi,\eta)$. Here, we
have taken $a=1$ and the irreversible entropy flux from the left
particle reservoir is $\alpha$ per entering particle. We expect
that systematically a net particle current flows through the
system, from the side with  higher chemical potential.
As we follow the path or trajectory $\eta(t)$ over some
time-interval $t\in [0,T]$
we can read the number of particles
that enter from the left (left time-integrated particle current)
and the number of particles that exit to the right.  These are of
course fluctuating currents, as the path $(\eta(t))$ is random with
a distribution obtained from the Markov dynamics.
For our purposes, the information on the
dynamics is summarized in the generator $L$, an $M\times M$ matrix
with elements $L(\eta,\xi) = k(\eta,\xi)$ for $\eta\neq \xi$ and
$L(\eta,\eta) =  - \sum_\xi k(\eta,\xi)$. Exactly because of the
nonequilibrium condition $\alpha\ne \alpha'$, the matrix $L$ need
not even be diagonalizable. For the final algorithm, that involves
a departure from the more standard Rayleigh-Schr\"odinger set-up,
as we might not have an orthogonal basis of eigenvectors.

\section{Path-space identity}

To understand the theoretical point of departure of the method, it
is useful to separate the time-antisymmetric part from the
time-symmetric part in the path-space distribution of the Markov
process.  Under the condition of local detailed balance (\ref{ldb}), the
time-antisymmetric part is directly related to the variable
entropy flux and the time-symmetric part measures the dynamical
activity, called {\em traffic}~\cite{TRAFFIC,TRAFFIC2} in the system.
Because of the normalization of the path-space
distribution these two sectors have related fluctuations, as we
will make explicit below in \eqref{cgf2}.  Specifically, we
consider ensembles of bonds $B=\{\eta\to\xi\}$ that all equally
contribute to the same time-integrated mesoscopic current
 $J_B=\int_0^T dJ_{B}(t)$.
In our example, we can take $B_1$ containing all transitions
getting one particle into the system and coming from the left
reservoir, and $B_2$ consisting of all transitions in which a
particle leaves the system to the right reservoir. We denote by
$-B$ the ensemble of reversed transitions, giving rise to an
instantaneous current $dJ_{-B} = -1$. We are then interested in
the various moments and correlations  between the $J_B$'s.  These
can be obtained from the cumulant-generating function
\begin{equation}\label{cgf}
g(\sigma) = \frac 1{T}\log\langle
\exp \sum_{B} \sigma_B J_{B}\rangle
\end{equation}
in the steady state of the system.
 For example the second (partial) derivatives with
respect to $\sigma_{B_1}, \sigma_{B_2}$ give a covariance between
two currents.  The point is now that the exponent in \eqref{cgf}
can be read as an excess entropy flux $\sum_{B} \sigma_B J_{B}$,
whose fluctuations are the same as that of a dynamical activity in
a Markov model with extra driving. To make that last point, we
imagine an extra driving by modifying the rates to
\[
L_\sigma(\eta,\xi)= k(\eta,\xi)\,e^{\sigma(\eta,\xi)}
\]
for some antisymmetric function $\sigma$, which, allowing some
abuse of notation, is $\sigma(\eta,\xi)=\pm\sigma_B$ if $(\eta,\xi)\in\pm B$.
One should check the local detailed balance condition (\ref{ldb}) 
to see that some extra
entropy flux is imposed. Both the original Markov process (with
generator $L$) and the modified one (with generator $L_\sigma$)
have a path-space distribution with action, respectively $A$ and
$A_\sigma$, and for which
\begin{equation}
\begin{split}
\exp\,[ T\,g(\sigma) ] &= \langle e^{-A + \sum_{B} \sigma_B J_{B}}\rangle_{\text{eq}}
\\
 &= \langle e^{-A_\sigma}\,e^{-A + A_\sigma + \sum_{B} \sigma_B
  J_{B}}\rangle_{\text{eq}}
\end{split}  
\end{equation}
with respect to some fixed equilibrium reference dynamics.
If we therefore arrange that the time-antisymmetric parts of $A$ and $A_\sigma$ differ exactly by the excess
entropy flux, we keep the excess in dynamical activity (time-symmetric parts):
\[
-A + A_\sigma + \sum_{B} \sigma_B J_{B} = \int_0^T V(\eta(t),\sigma)\,dt
\]
for a particular function $V$ on the state space, that also depends
on the $\sigma$. That function $V(\eta)$ essentially measures the
difference in escape rates from $\eta$ for the original process
with respect to the one modified by the $\sigma$'s. 
The conclusion is a general path-space identity
\begin{equation}\label{cgf2}
g(\sigma) = \frac 1{T}\log\langle
\exp \int_0^T V(\eta(t),\sigma)\,dt \rangle_\sigma
\end{equation}
where the last expectation is for the modified steady state.  The
formula \eqref{cgf2} is more ready for asymptotic evaluation as
$T\uparrow +\infty$
since $V$ is a multiplication operator. The
algorithm must then give a systematic expansion in the $\sigma_B$
of the largest eigenvalue (in the sense of its real part) of the
matrix $\caL = L_\sigma + V(\cdot,\sigma) = L + \caR = L + \sum_B \sum_{n\ge 1} (\sigma_B)^{n} \caL_B^{(n)}$, where 
$\caR$ is the matrix with non-zero elements 
$[e^{\pm\sigma_B}-1]k(\eta,\xi)$ only in ensembles $\pm B$'s, where
the rates of $L_\sigma$ are different from the rates of $L$. 
The details of this expansion are in~\cite{bmn08}.
We just briefly mention that this expansion does not have the
advantage of dealing with symmetric matrices, 
as in the original Rayleigh-Schr\"odinger 
method for quantum mechanical operators.
The solution of the problem goes via the use of {\em resolvents}.
Interestingly, 
the final results include the group inverse $G$ of the generator $L$
as a main actor~\cite{mey1,mey2}. 
This matrix contains all the informations needed to
compute the dynamical quantities of interest, like
the stationary distribution $\BRA{\rho}$ and the cumulants.
For example, the variance of a current takes the form
$C_{BB} =\BRA{\rho} \caL_B^{(2)} \KET{1} - 
\BRA{\rho} \caL_B^{(1)}G \caL_B^{(1)} \KET{1}$ (by $\KET{1}$ we mean 
a vector of $1$'s).

\section{Illustration}

\begin{figure}[!t]
  \includegraphics[width=.9\textwidth]{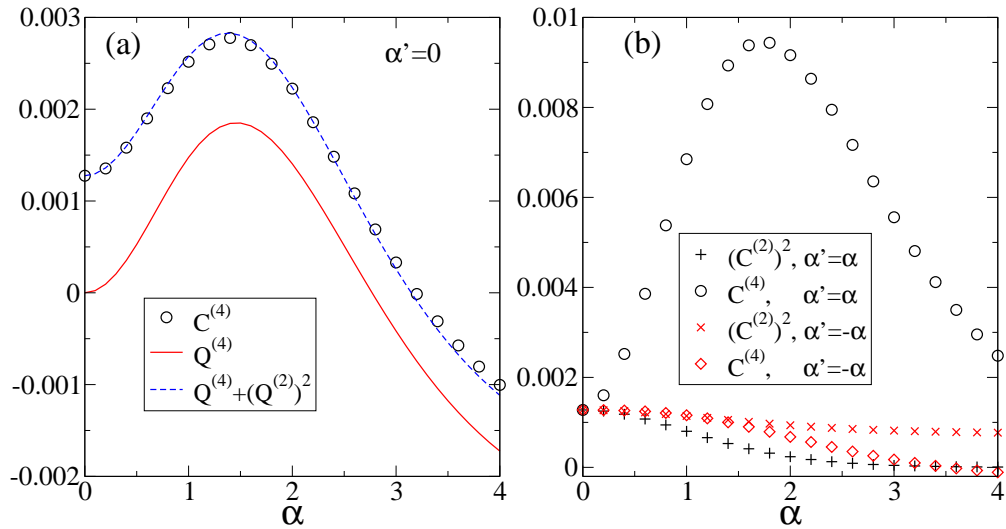}
  \caption{(a) Fourth cumulant ($C^{(4)}$)
of the open SSEP as a function of $\alpha$, and the theoretical
value ($Q^{(4)}$) by Derrida and coworkers, for $N=14$ and $\alpha'=0$.
The theoretical $Q^{(4)}+(Q^{(2)})^2$ is also shown.
(b)~$(C^{(2)})^2$ and $C^{(4)}$ vs $\alpha$ for $N=14$,  with $\alpha'=\alpha$ 
and  $\alpha'=-\alpha$ (see legend).}
\label{fig}
\end{figure}

We end with a discussion of some results for the open SSEP introduced above.
For the SSEP on a ring, cumulants $Q^{(n)}$ of the current have been computed
theoretically~\cite{derrida08}.
Formulas for the mean current and its variance 
are also available for the boundary-driven SSEP,
while formulas for cumulants of order $n\ge 3$ are known up to order
$1/N$~\cite{derex,derrida}.  
Being one-dimensional and finite,
there is only one relevant current in the system, which we identify with
the entrance of a particle from the left reservoir.
Our data, for system sizes up to $N=14$, at equilibrium  
with half-filling
($\alpha=\alpha'=0$)\footnote{
The chemical potentials $\alpha$ and $\alpha'$ correspond
to reservoirs with particle density
$\rho_a=\exp(\alpha/2)/[\exp(\alpha/2)+\exp(-\alpha/2))$ and
$\rho_b=\exp(\alpha'/2)/[\exp(\alpha'/2)+\exp(-\alpha'/2))$
in~\cite{derex,derrida}.}
perfectly agree with the theoretical value:  
$C^{(2)}=Q^{(2)} = (2 N)^{-1}$.
In the same conditions, for the fourth cumulant we observe
\begin{equation}
C^{(4)} = (C^{(2)})^2 = \frac{1}{(2N)^2}
\label{eq:C4}
\end{equation}
for $N\le 14$, while to order $1/N$ one has $Q^{(4)}=0$~\cite{derex,derrida}.
The relation (\ref{eq:C4}) is slightly
different from the one found for the SSEP at equilibrium on a ring,
where the asymptotic values of the second and fourth cumulants of the current
are related by
$Q^{(4)} = \frac{1}{2} (Q^{(2)})^2$~\cite{derrida08}.
It turns out that the addition of a term $=(Q^{(2)})^2$ to $Q^{(4)}$
yields a good approximation of $C^{(4)}$ for finite $N$
also out of equilibrium, see the Fig.~\ref{fig}(a).
Finally, we have tested that the 
relation (\ref{eq:C4}) is not valid for $\alpha=\alpha'\ne 0$ 
(no half-filling) or for
$\alpha=-\alpha'\ne 0$ (a case of half-filling out of equilibrium), see 
Fig.~\ref{fig}(b).

\begin{theacknowledgments}
M.~B.~acknowledges financial support from K.~U.~Leuven grant OT/07/034A.
C.~M.~benefits from the Belgian Interuniversity Attraction Poles Programme
P6/02.
K.~N.~thanks Tom\'a\v{s} Novotn\'y for fruitful discussions
and acknowledges the support from the project AVOZ10100520 
in the Academy of Sciences of the Czech Republic and from the 
Grant Agency of the Czech Republic (Grant no.~202/07/J051).
\end{theacknowledgments}



\bibliographystyle{aipproc}   


\end{document}